\documentclass[aps,prl,twocolumn,showpacs,superscriptaddress,groupedaddress]{revtex4}  
\usepackage{graphicx}  
\usepackage{epsfig}
\usepackage{subfigure}
\usepackage{dcolumn}   
\usepackage{bm}        
\usepackage{amssymb}   
\def\e{\begin{equation}}
\def\f{\end{equation}}
\def\%#1{\mbox{\boldmath $#1$}}
\def\=#1{\overline{\overline #1}}
\def\*#1{\overline{\overline{\overline #1}}}

\def\-#1{{\bf #1}}

\def\va{\varepsilon}

\def\.{\cdot}

\def\##1{{\bf#1\mit}}

\def\am{\left(\begin{array}{c}}
\def\amm{\left(\begin{array}{cc}}
\def\a{\end{array}\right)}

\begin{document}



\title{Huge local field enhancement in perfect plasmonic absorbers}

\author{M. Albooyeh}

\author{C. R. Simovski}

\affiliation{Department of Radio Science and Engineering / SMARAD
Centre of Excellence,\\ Aalto University, P.O. Box 13000, FI-00076
Aalto, Finland}

\date{\today }

\begin{abstract}

In this Letter we theoretically study the possibility of total
power absorption of light in a planar grid modelled as an
effective sheet with zero optical thickness. The key prerequisite
of this effect is the simultaneous presence of both resonant
electric and magnetic modes in the structure. We show that the
needed level of the magnetic mode is achievable using the effect
of substrate-induced bianisotropy which also allows the huge local
field enhancement at the same wavelength where the maximal
absorption holds.
\end{abstract}

\maketitle

Total absorption (TA) of light at a certain wavelength in a lossy
layer located on a substrate implies simultaneous suppression of
both reflection of the incident wave and transmission of this wave
into the substrate. This effect has been applied in optical
sensing and energy harvesting devices~\cite{Luo1, Yu1, Sch1, Der1,
Eme1, Polman}. However, in the majority known optical absorbers
the TA regime is not achievable and can be only more or less
approached \cite{Azzam}. In the microwave and mm wave ranges the
TA was engineered exactly in the so-called Dallenbach absorber
(see e.g. in \cite{Vin}) which comprises a metal substrate. In the
optical range the TA was obtained in the so-called coherent
perfect absorber using two incident waves \cite{Sci}. In both
these cases the absorber was a bulk lossy layer and the
destructive wave interference in presence of two reflecting
boundaries was used. Recently, the TA has been predicted for
so-called perfect plasmonic absorbers (PPA) \cite{PPA1,PPA2,PPA3}.
PPA is a very thin and optically dense planar grid of plasmonic
nanoparticles (nanodisks or nanopatches) separated by an optically
thin dielectric gap from a metal substrate. Planar grids of
resonant particles can be homogenized using the surface averaging
of their microscopic responses (see e.g.~\cite{Hol2, Sim1}). Such
homogenized resonant grids presented as infinitesimally thin
sheets of surface polarization were called metasurfaces or
metafilms. The mechanism of the TA in known PPAs is still the
destructive wave interference. The wave reflected from the metal
substrate cancels out with that reflected from the metafilm
\cite{PPA2,PPA3}. This cancellation becomes possible due to the
resonant properties of the plasmonic grid.

In the present Letter we suggest a novel type of PPA with another
mechanism of the TA and with an additional functionality. In our
PPA the transmittance into the substrate is prevented even if the
substrate is transparent and the reflectance is prevented in the
absence of the second reflecting boundary. The TA is achieved due
to the combination of the electric and magnetic modes in the
structure. Additionally, our PPAs possess huge local field
enhancement (LFE) in strongly subwavelength volumes. The effect of
LFE is very important in numerous applications and has been
recently widely studied (see e.g.~\cite{Kim1, Tam1, Wat1, Mic1,
Led1, Gri1, Rot1, Chall1}). The combination of TA and LFE is very
advantageous, e.g. for the surface enhanced Raman scattering
(SERS). In many of SERS schemes the LFE holds in a targeted
spatial region (e.g. at the nanopatterned surface of silver or
gold) and results in the giant enhancement of the Raman radiation
from this region \cite{Raman,Raman1}. Both reflected and
transmitted waves are parasitic signals whose presence hinders the
detection of the Raman radiation from the targeted spatial region
\cite{Raman,Raman1}. Thus, for prospective SERS schemes it is
advantageous to use PPAs (if they possess the LFE effect).

\begin{figure}[ht]
\subfigure[]{\includegraphics[width=0.45\linewidth]{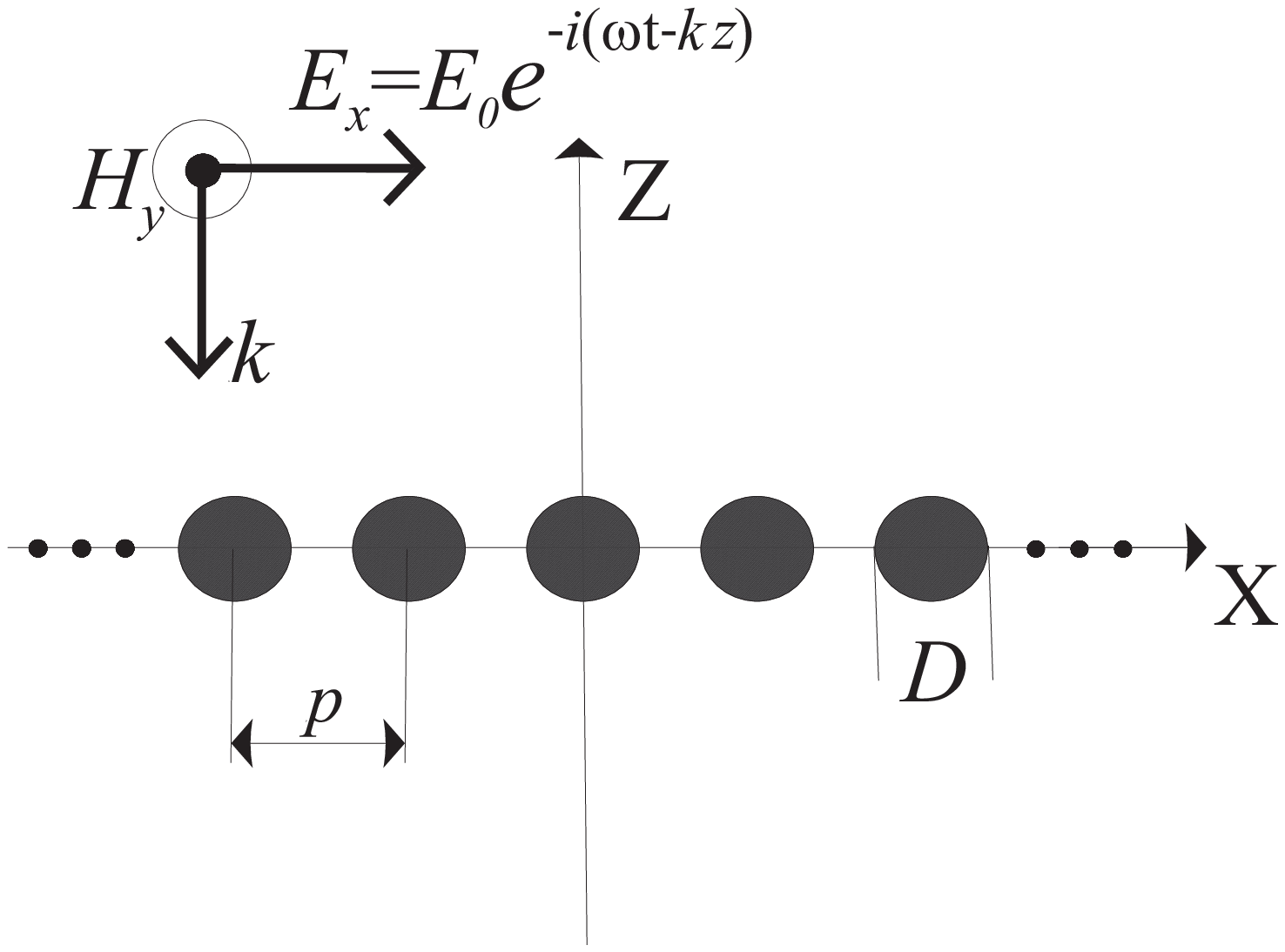}}
\subfigure[]{\includegraphics[width=0.4\linewidth]{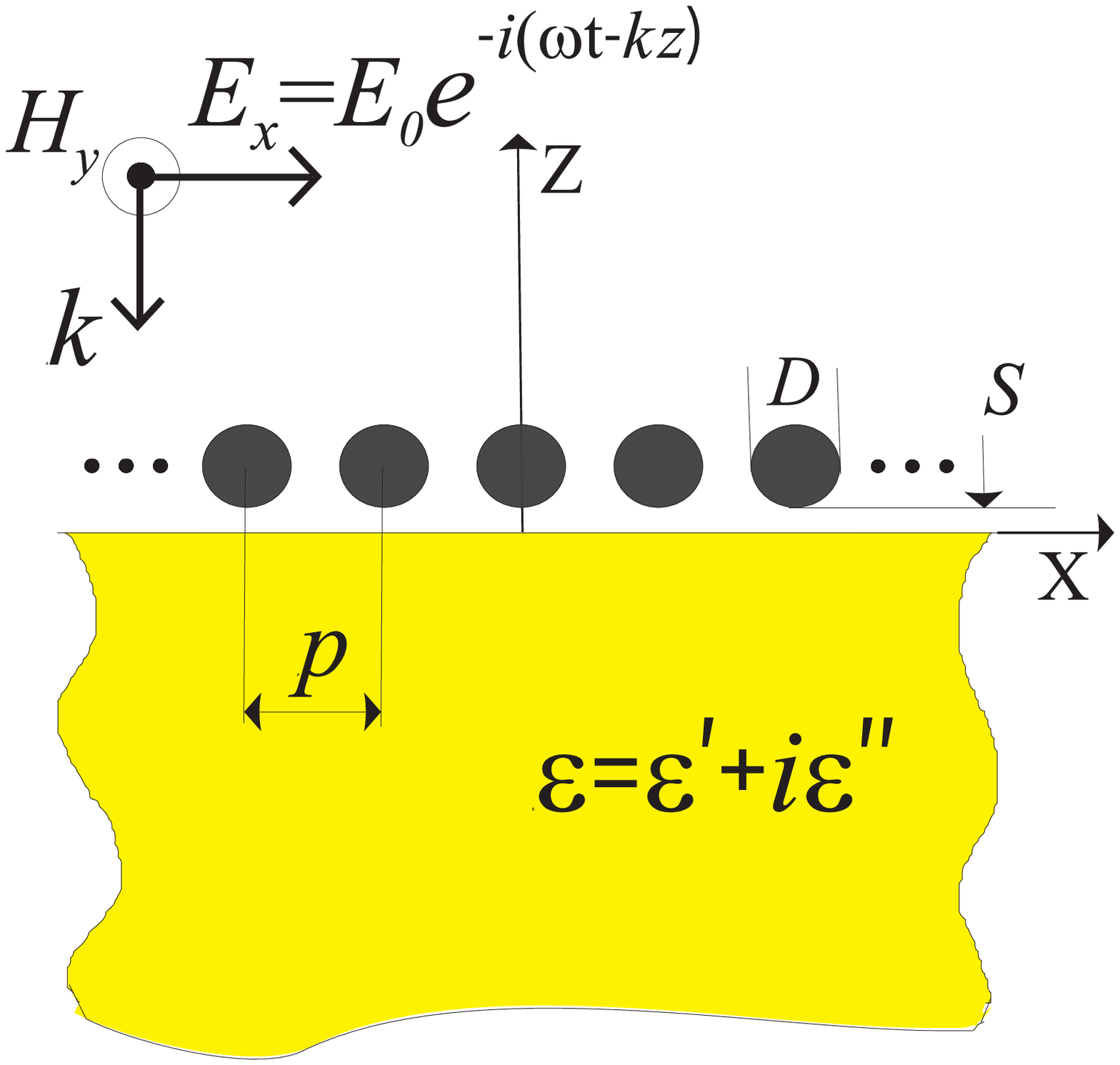}}
\caption{(Color online) (a) -- Horizontally isotropic scatterers
with both electric and magnetic responses (e.g. magneto-dielectric
spheres experiencing the magnetic and electric Mie resonances at
microwaves). (b) -- Horizontally isotropic grid of plasmonic
particles located over a semiconductor substrate.}
\label{fig:figSchem1}
\end{figure}

First, let us show that it is possible to achieve the TA in a
metafilm without any substrate. Consider an optically dense planar
array of particles possessing both electric and magnetic
polarizabilities and located in free space. Alternatively, we can
have in mind one electric dipole and one magnetic dipole per unit
cell located in the grid plane. This planar array can be modelled
as a metafim with both electric and magnetic surface
polarizations~\cite{Hol2, Sim1}. Let us show that the TA in the
metafilm is  achievable with realizable parameters describing its
electromagnetic response. For simplicity the array is assumed to
be isotropic in the $XY$-plane and a plane wave impinging normally
to the array plane, see Fig.~\ref{fig:figSchem1} (a). Then
according to ~\cite{Hol2, Mo1} one can describe the response of the
effective sheet by two surface susceptibilities $\alpha^{ee}_{xx}$
and $\alpha^{mm}_{yy}$. The relationship between the electric and
magnetic surface polarizations and the mean values of the electric
and magnetic fields at the grid plane can be written as
\cite{Hol2, Mo1}: \e
\begin{array}{ccc}
 {{\cal P}_x} =  \alpha^{ee}_{xx}<E_x>,  \quad {{\cal M}_y} = \alpha^{mm}_{yy}<H_y>.
\end{array}
\label{eq:defs1}\f Here $<\-E>$ and $<\-H>$ are defined as the averaged
values of electric and magnetic fields taken on two sides of the
effective sheet, ${\cal P}_x$ and ${\cal M}_y$ are the electric
and magnetic surface polarizations, respectively. Using the
boundary conditions for such metafilms derived in \cite{Hol2, Mo1}, we
obtain from ({\ref{eq:defs1}}): \e R=-{{X}\over {1+X}}+{{Y}\over {1+Y}},
\quad T={1-{{X}\over {1+X}}-{{Y}\over {1+Y}}}.\label{eq:RefTrS1}\f Here
$R$ and $T$ are complex reflection and transmission coefficients,
respectively, and it is denoted: $X=-{i\over 2}{k_0\over
{\va_0}}\alpha^{ee}_{xx}$, $Y=-{i\over 2}{k_0\over
{\mu_0}}\alpha^{mm}_{yy}$, where $k_0$ is the free space wave
number (${\mu_0}$ and ${\va_0}$ are free space permeability and
permittivity, respectively). From~(\ref{eq:RefTrS1}) it is easy to see
that $R=T=0$ if electric and magnetic susceptibilities are,
respectively, equal to:\e
  \alpha^{ee}_{xx}={2i{\va_0\over k_0}}, ~~(a), \quad \alpha^{mm}_{yy}={2i{{\mu_0}\over k_0}}. ~~(b) \\
\label{eq:twoe}\f



Both required electric and magnetic polarizabilities are purely
imaginary that is achievable if the electric and magnetic
resonances of our particles coincide. This is definitely a
realizable requirement i.e. the regime when the absorption
coefficient $A\equiv 1-|R|^2-|T|^2$ equals to unity is possible.
The key prerequisite is the resonant magnetic susceptibility~(3-b)
created in addition to the electric one~(3-a). This is not
problematic at microwaves, however for the visible frequency range
becomes challenging. Therefore further we show how to engineer the
surface magnetic polarization for a metafilm without any magnetic
susceptibility in this range. Moreover we show how to combine the
TA together with the LFE obtaining this way a new functionality of PPA
layers.

The proposed PPA depicted in Fig.~\ref{fig:figSchem1} (b)
comprises an optically dense planar grid (e.g. a square array with
the period $p\ll\lambda$) of plasmonic nanospheres with the
optically small diameter $D$ located over the surface of a
strongly refracting half-space at a very small distance $s$ from
it. In previously known PPAs plasmonic nanoparticles were thin
tablets and the plasmonic grid formed the effective plane of
reflection clearly distinct from a metal interface. Thus, the
reflection in previously known PPAs was prevented due to the wave
interference \cite{PPA2} whereas the transmittance was prevented
due to the total reflectance of the metal substrate. In the
present case the distance $s$ between the lower edges of
nanospheres and the semiconductor surface is much smaller than $D$
and the geometry does not contain two effective reflection planes.
The only reflection plane is the interface to which we refer our
metafilm. The resonant magnetic polarization in our metafilm
arises due to the so-called substrate-induced bianisotropy which
was theoretically revealed and studied in \cite{Mo1} for arrays of
resonant electric dipoles located on top of a half-space with high
refractive index. For the visible range this implies a
semiconductor substrate. In our case both reflection and
transmission are prevented to the balance of the effective
electric and magnetic polarizations of a metafilm. However, the
most important effect is the regime of LFE i.e. the new suggested
functionality of our PPA.

\begin{table}[t!]
\caption{\label{tab:Dims1}Dimensions of the PPA and characteristic
parameters of materials at the wavelength $\lambda_{MPA}$=361.9
nm.}
\begin{ruledtabular}
\begin{tabular}{cccc}
$D$ (nm)& $p$ (nm)& $s$ (nm)& $\va = \va'+i\va''$ \\
\hline
$60$& $120$ & $4$ & $12.8+i1.0$ (a-Si substrate)\\
& & & $-1.892+i0.036$ (Silver spheres)\\
\end{tabular}
\end{ruledtabular}
\end{table}

\begin{figure}
\includegraphics[width=0.3\textwidth,height=0.28\textwidth]{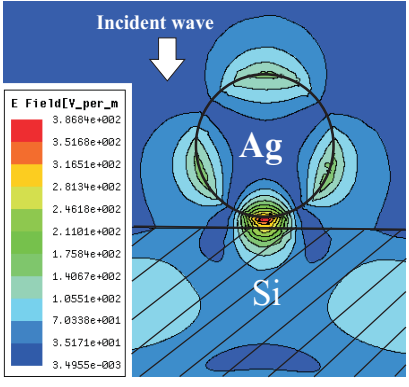}
\caption{\label{fig:fignew} (Color online) Color map of the local
field amplitude normalized to that of the incident wave at the
wavelength of maximal power absorption.}
\end{figure}

In this paper we consider the normally incident plane wave and a
free space gap $s$. This is done for simplicity and to stress that
the wave interference does not play any role. The same
functionality can be shown for obliquely incident waves and for
nanospheres located on a 1-3 nanometer thick intermediate layer
with low refractive index. Although small plasmonic nanospheres
possess only the electric dipole response, the presence of the
high-contrast substrate leads to the excitation of the magnetic
mode~\cite{Mo1}. This fact is illustrated by Fig.~\ref{fig:fignew}
which also shows the LFE in the proposed scheme of the PPA.
Fig.~\ref{fig:fignew} is the color map of the electric field
amplitude in the central cross section of a unit cell. This field
distribution was simulated at the wavelength $\lambda_{MPA}$ of
maximal power absorption (MPA) achievable for this structure ($A=0.99$).
Dimensions of the structure and parameters of the used materials
at $\lambda_{MPA}$ are summarized in Table~\ref{tab:Dims1}, see
also Fig.~\ref{fig:figSchem1} (b). The complex permittivities of
Ag and a-Si were taken from \cite{El-K} and \cite{Si},
respectively. Since $\lambda_{MPA}$ is within the range of the
plasmon resonance of the nanosphere it acquires a large dipole
moment. In accordance with the quasi-static image principle
calculating the near field one can replace the substrate by an
image of the nanosphere. The local field is enhanced mainly in the
gap between the nanosphere and the substrate (see
Fig.~\ref{fig:fignew}). This is similar to the operation of a
nanoantenna \cite{Tam1}. This is not a usual situation since the hot
spot in the gap arises for the horizontal polarization of the
incident field. However a similar mode has been known for
non-symmetric dimers of plasmonic spheres \cite{Klimov}. In this
dimer the mode analogous to our one is characterized by a strong
local polarization of both spheres near their internal edges. In
our case the role of the second sphere is played by the image of
the first one created by the substrate. To obtain this regime we
need a strong optical contrast between two half-spaces and a
sufficient curvature of the nanoparticle.

Fig.~\ref{fig:fignew} also illustrates the substrate-induced
bianisotropy. Since the gap $s$ is very small a noticeable part of
the hot spot turns out to be located inside the semiconductor.
This local polarization of the substrate is concentrated in a
nanovolume at a noticeable distance from the effective dipole of
the nanosphere. One can check that at $\lambda_{MPA}$ these two
effective dipoles (they are both horizontally directed) have
opposite phases. The unit cell of the proposed PPA is then
equivalent to a pair of electric and magnetic dipoles. Due to the
optical smallness of the cell both these dipoles can be referred
to the interface. After the homogenization the structure becomes
an effective sheet of resonant electric and magnetic surface
polarizations. Whereas the LFE in the PPA results from the
curvature of the nanoparticle, the TA can be explained as we have
done above -- using the homogenization model.

Using the same model as in~\cite{Mo1} three effective tangential
susceptibilities per unit area of the metafilm can be introduced
by relations:\e
\begin{array}{ccc}
 {{\cal P}_x} =  \alpha^{ee}_{xx}<E_x>+ \alpha^{em}_{xy}<H_y>,  \quad {{\cal M}_y} = \alpha^{me}_{yx}<E_x>.
\end{array}
\label{eq:defs}\f Here $\alpha^{em}_{xy}$ and $\alpha^{me}_{yx}$ are the
electro-magnetic and magneto-electric surface susceptibilities,
respectively. Due to the reciprocity we have
$\alpha^{em}_{xy}=-\alpha^{me}_{yx}$, and only two
susceptibilities remain in (\ref{eq:defs}). Using boundary conditions
derived in~\cite{Mo1} we relate the complex reflection and
transmission coefficients $R$ and $T$ with these susceptibilities
and obtain the retrieval formulas:\e
\alpha^{ee}_{xx}={i{4\va_0\over k_0}}{{1-R^2-nT^2}\over
{(1+R+T)}^2}, \quad \alpha^{em}_{xy}={i{2\sqrt{\va_0 \mu_0}\over
k_0}}{{1+R-T}\over {1+R+T}} \label{eq:Susc}\f

The solution corresponding to the total absorption when $R=T=0$
and $A=1$ is evident:\e \alpha^{ee}_{xx}={4i{\va_0\over k_0}},
\quad \alpha^{em}_{xy}={2i{\sqrt{\va_0 \mu_0}\over k_0}}.
\label{eq:RTZC}\f Imaginary electric susceptibility in (\ref{eq:RTZC}) is
achievable exactly and corresponds to the maximum of electric
losses in the metafilm. Imaginary magneto-electric susceptibility,
on the contrary, corresponds to the absence of magneto-electric
losses \cite{Serd}. Strictly speaking the TA regime corresponding
conditions (\ref{eq:RTZC}) is an idealization: one can hardly create a
metafilm with lossless electro-magnetic (magneto-electric)
response. However, one can approach the target conditions (\ref{eq:RTZC})
very closely. We refer to this regime as to that of maximal power
absorption.

\begin{figure}
\includegraphics[width=0.4\textwidth,height=0.3\textwidth]{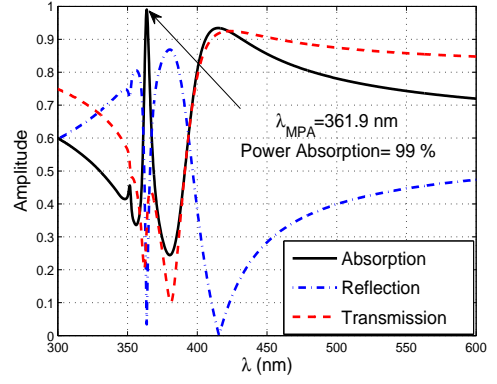}
\caption{\label{fig:figRTA} (Color online) Coefficients of
reflection (amplitude), transmission (amplitude), and absorption
(power) of the normally incident plane wave in the proposed
plasmonic structure.}
\end{figure}

\begin{table}
\caption{\label{tab:SurfaceS}Desired and achieved surface
susceptibilities of the metafilm with $s=4$ nm at the wavelength
$\lambda_{MPA}=$361.9 nm.}
\begin{ruledtabular}
\begin{tabular}{ccc}
 susceptibility &$Desired$&$Obtained$\\
\hline

$\alpha^{ee}_{xx} ~(F)$& $2.04\times10^{-18}(0+i1)$ & $2.97\times10^{-18}(0.20+i1)$ \\
$\alpha^{em}_{xy} ~(s)$& $3.84\times10^{-16}(0+i1)$ & $4.52\times10^{-16}(0.26+i1)$ \\
\end{tabular}
\end{ruledtabular}
\end{table}

\begin{figure}
\includegraphics[width=0.3\textwidth,height=0.25\textwidth]{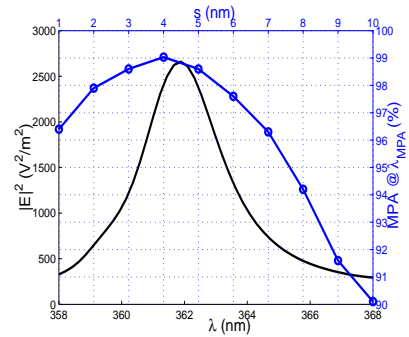}
\caption{\label{fig:figLFE} (Color online) Black line: the square
power of the local electric field amplitude normalized to that of
the incident wave versus wavelength for $s=4$ nm. The local field
is calculated at the center of the gap. Blue line with circles:
the absorbtion coefficient in percent as a function of $s$
calculated at wavelengths $\lambda_{MPA}(s)$.}
\end{figure}

In our simulations we used both options of the HFSS package (the
incident plane wave and the wave port) thoroughly checking the
convergence. Moreover, we have reproduced the same results using
the CST Studio software. Reflection, transmission, and absorption coefficients of this PPA versus
frequency are plotted in Fig.~\ref{fig:figRTA}. Using simulated
$R$ and $T$ coefficients (amplitudes and phases) and substituting
these data into~(\ref{eq:Susc}) we have retrieved the corresponding
electric and electro-magnetic susceptibilities. The target
susceptibilities~(\ref{eq:RTZC}) and those retrieved for the proposed PPA
at the wavelength $\lambda_{MPA}$ are presented for comparison in
Table~\ref{tab:SurfaceS}. For spheres with $D=60$ nm over the a-Si
substrate this disagreement between ideal and retrieved
susceptibilities is minimal over all possible $\lambda$ and $s$.
Further optimization is possible varying $D$ and $p$.

In Fig.~\ref{fig:figLFE} the line with circles shows the
dependency $A(s)$ calculated for every $s$ at the corresponding
$\lambda_{MPA}$ (for $s$ increasing from 1 to 10 nm the wavelength
of MPA $\lambda_{MPA}$ decreases from 381 to 358 nm). The
resonance of the local field intensity normalized to that of the
incident wave calculated at the brightest point of the hot spot is
illustrated by the black line without circles in
Fig.~\ref{fig:figLFE} for the optimal height $s=4$ nm. The effect
of the maximal LFE holds exactly at the same wavelength as that of
MPA $\lambda_{LFE}=\lambda_{MPA}\approx 362$ nm. The same
observation refers to other values of $s$.

In this Letter we have theoretically demonstrated a possibility of
almost total absorption in an optically dense planar grid of
plasmonic nanoparticles over a transparent substrate. We have
shown that this absorption does not result from the destructive
wave interference as in previously known PPAs. The effect holds
due to the balance of the electric and magnetic polarizations
whereas the last one results from the substrate-induced
bianisotropy. Moreover, this absorption is accompanied by a huge
local field enhancement. This new functionality is granted by an
effect similar to the LFE in plasmonic nanoantennas. We have
checked by numerous simulations that this huge LFE does not hold
for arrays of nanotablets, semi-spheres and other particles
without curvature of the lower part. We have found that the larger
is this curvature the stronger is the LFE. Though in the present
paper the gap between the Ag nanospheres and the a-Si substrate is
assumed to be filled with free space, this is not critical.
Similar results can be obtained for spheres located in a
dielectric medium over another semiconductor substrate.
Alternatively, nanoparticles can be located in free space on a
very thin dielectric layer. Local field intensity enhancement
$(2-3)\cdot 10^3$ that we have revealed above (which is typical
for many SERS schemes with nanoantennas) can be once more
increased in arrays of touching nanopspheres due to collective
effects \cite{Pendry,Shalaev}. This modification of the PPA can be
probably done without the damage for the MPA regime.


Finally, we want to thank Prof. Sergei Tretyakov and Dr. Pavel
Belov for clarifying and fruitful discussions.


\begin{thebibliography}{99}




\bibitem{Luo1}
C. Luo, A. Narayanaswamy, G. Chen, and J. D. Joannopoulos,
Phys. Rev. Lett. {\bf 93}, 213905 (2004).

\bibitem{Yu1}
Z. Yu, G. Veronis, S. Fan, and M. L. Brongersma, Appl. Phys.
Lett. {\bf 89}, 151116 (2006).

 \bibitem{Sch1}
D. M. Schaadt, B. Feng, and E. T. Yu, Appl. Phys. Lett. {\bf 86},
063106 (2005).


 \bibitem{Der1}
D. Derkacs, S. H. Lim, P. Matheu, W. Mar, and E. T. Yu, Appl.
Phys. Lett. {\bf 89}, 093103 (2006).

\bibitem{Eme1}
K. Emery, Semicond. Sci. Technol. {\bf 18}, 228 (2003).

\bibitem{Polman}
H. A. Atwater and A. Polman, {Nature Mat.} \textbf{9}, 205 (2010).

\bibitem{Azzam}
R. M. A. Azzam, E. Bu-Habib, J. Casset, G. Chassaing, and P.
Gravier, Appl. Optics \textbf{26}, 719 (1987).


\bibitem{Vin}
K. J. Vinoy, R. M. Jha, \emph{Radar Absorbing Materials: From
Theory to Design and Characterization}, Kluwer: Boston, USA, 1999.

\bibitem{Sci}
W. J. Wan, Y. D. Chong, L. Ge, H. Noh, A. D. Stone, and H. Cao,
Science \textbf{331}, 889 (2011).


\bibitem{PPA1}
N. Liu, M. Mesch, T.s Weiss, M. Hentschel, and H. Giessen, Nano
Lett. \textbf{10}, 2342 (2010).

\bibitem{PPA2}
M. Pu, C. Hu, M. Wang, C. Huang, Z. Zhao, C. Wang, Q. Feng, and X.
Luo, Opt. Expr. \textbf{19}, 17413 (2011).

\bibitem{PPA3}
M. K. Hedayati {\sl et al.}
Adv. Mat. \textbf{23}, 5410
(2011).


 \bibitem{Hol2}
C.L. Holloway, M.A. Mohamed, E.F. Kuester, and A. Dienstfrey, IEEE
Trans. Electromagn. Compat. {\bf 47}, 853 (2005).

 \bibitem{Sim1}
C. R. Simovski, J. Opt. {\bf 13}, 013001 (2011).

  \bibitem{Kim1}
S. Kim {\sl et al.}, Nature (London) {\bf 453}, 757 (2008).

  \bibitem{Tam1}
T. H. Taminiau {\sl et al.}, Nat. Photon. {\bf 2}, 234 (2008).

  \bibitem{Wat1}
K. Watanabe {\sl et al.}, Chem. Rev. {\bf 106}, 4301 (2006).

  \bibitem{Mic1}
X. Michalet {\sl et al.}, Science {\bf 307}, 538 (2005).

  \bibitem{Led1}
K.W. D. Ledingham, P. McKenna, and R. P. Singhal,
Science {\bf 300}, 1107 (2003).

  \bibitem{Gri1}
D. G. Grier, Nature (London) {\bf 424}, 810 (2003).

  \bibitem{Rot1}
B. Rothenhausler and W. Knoll, Nature (London) {\bf 332}, 615
(1988).

  \bibitem{Chall1}
W. A. Challenger {\sl et al.}, Nat. Photon. {\bf 3}, 220 (2009).

\bibitem{Raman}
E. C. Le Ru and P. G. Etchegoin, \emph{Principles of
Surface-Enhanced Raman Spectroscopy and Related Plasmonic
Effects}, Elsevier: Oxford, UK, 2009.


\bibitem{Raman1}
\emph{Surface-Enhanced Raman Scattering, Physics and
Applications}, K. Kneipp, M. Moskovits, and H. Kneipp, Eds.,
Springer: Berlin-- Heidelberg--New York, 2006.




 \bibitem{Mo1}
M. Albooyeh and C. Simovski, J. Opt. {\bf 13}, 105102 (2011).


\bibitem{El-K}
I. El-Kady, M. M. Sigalas, R. Biswas, K. M. Ho, and C. M. Soukoulis, Phys. Rev. B {\bf 62}, 15299 (2000).


\bibitem{Si}
A. S. Ferlauto, G. M. Ferreira, J. M. Pearce, C. R. Wronski, R. W.
Collins, X. Deng, and G. Ganguly, J. Applied Phys. \textbf{92},
2424 (2002).

\bibitem{Klimov}
V. V. Klimov and D. V. Guzatov, Phys.Rev. B \textbf{75},024303
(2007).

  \bibitem{Serd}
A. Serdyukov, I. Semchenko, S. Tretyakov, and A. Sihvola,
\emph{Electromagnetics of Bi-anisotropic Materials: Theory and
Applications}, Gordon and Breach Science: Amsterdam, The
Netherlands, 2001.

  \bibitem{Pendry}
F.J. Garcia-Vidal and J.B. Pendry, \emph{Phys. Rev. Lett.}
\textbf{77}, 1163 (1996).

  \bibitem{Shalaev}
D. A. Genov, A. K. Sarychev, V. M. Shalaev, and A. Wei, Nano Lett.
\textbf{4}, 153 (2004).


\end{thebibliography}
\end{document}